\documentstyle[11pt,paspconf,epsf]{article}

\begin{document}

\title{ HST snapshot imaging of BL Lac objects}

\author{R. Falomo }
\affil{Osservatorio Astronomico di Padova, Italy }
\author{C. Megan Urry \& Riccardo Scarpa } 
\affil{Space Telescope Science Institute}
\author{Joseph E. Pesce }
\affil{Department of Astronomy, Pennsylvania State University, USA}
\author{Aldo Treves } 
\affil{University of Como, Italy}

\begin{abstract}
Snapshot images of $\sim$ 100 BL Lac objects were obtained with 
WFPC2 on HST. Sources from various samples, in the
redshift range 0.05 to 1.2, were observed and  61 resolved (51 with
known $z$). 
 The high resolution and homogeneity of the
images allow us to address the properties of the immediate
environments of BL Lacs with unprecedented capability. 
Host galaxies of BL Lacs are luminous ellipticals (on average 1 mag
brighter than L$^*$) with no or little disturbed morphology. 
 The nucleus, that is always well centered
onto the galaxy, contributes in the optical (R band) to about half of
the total luminosity of the object (range of the contribution from 0.1
to 10).  The undisturbed morphology suggests that the nuclear activity has
marginal effect on the overall properties of the hosts. Nonetheless
several examples of close companions have been detected.
 The luminosity distribution of host galaxies is  compared with that of 
 a large sample of FR-I radio galaxies.
\end{abstract}



\section{Introduction}
Imaging of the galaxies hosting active galactic nuclei (AGN) is an
important tool for the understanding of the activity phenomenon
itself.  This study is often hampered by the presence of the bright
nucleus that swamps the light from the host galaxy.  This is clearly a
more serious problem for distant objects where the Point Spread
Function (PSF) of the nuclear source becomes comparable with the size
of the host galaxy.
 
The use of HST data has improved the capability to investigate the
galaxies hosting nuclear activity and in fact a number of specific
studies of nearby and intermediate redshift AGN have been pursued (see
e.g. Disney et al. 1995 ; Bahcall et al. 1997; Hooper, Impey \& Foltz
1997; Malkan Gorjian and Raymond 1998).

BL Lac objects are radio loud AGN seen closely along the jet emission
(see e.g. the review by Urry and Padovani 1995). 
The beaming properties of the jets
suggest that low-luminosity radio galaxies are the corresponding
mis-aligned population. Observations of the host galaxies are a direct
probe of this unification hypothesis.

We report here some results on $\sim$ 100  BL Lac objects imaged with HST
during a snapshot program ( see Urry et al
1999b for full discussion).  Previous imaging studies of BL
Lacs using HST (GO programs) have been performed only for a limited 
 number of sources (see Falomo et al 1997,
Januzzi et al 1997 and Urry et al 1999a). 

Taking advantage of high spatial resolution, homogeneity of data
quality and analysis and sizeable data set, we are able to progress
substantially  on the luminosity distribution and
morphological type of host galaxies, on the issue of the centering of
the nucleus on the main body of the galaxy (microlensing) and on
relevance of interactions with close companions.

\section{Observations and data analysis}

We have obtained HST snapshot images for 97 BL Lac objects from six
 samples (see Urry et al 1999b).  Images of the targets were collected
 using the WFPC2 camera and the F702W (roughly R ) filter. For each
 source we obtained typically 2-3 images with exposure time ranging
 from 2 to 10 minutes. The objects were centered in the PC CCD.  After
 data reduction following the standard {\it HST} pipeline processing
 we have combined individual frames of the same object and used
 photometric calibration converted to Cousins R band as described in
 Scarpa et al (1999).

In order to understand if the sources are resolved and to derive the
properties of the host galaxy a reliable model for the PSF is needed.
The Tiny Tim model (Krist 1995) gives an excellent representation of
the PSF shape within $\sim 2-3$~arcsec from the center. Outside this
range, there is an extra emission due to the scattered light that
affects the PSF wings.  To account for this we used observations of
saturated stars from the HST archive and modeled the extra emission
with an exponential component (see Scarpa et al 1999 for more
details).

The properties of the BL~Lac host galaxies were derived both from
azimuthally averaged radial profiles and from two-dimensional surface
photometry (isophote fitting) when enough signal was present.  We
fitted the luminosity profile as a function of radius with models
consisting of a nuclear unresolved source plus a galaxy (de
Vaucouleurs $r^{1/4}$ law for an elliptical or exponential law for a
disc), convolved with the PSF.  Surface photometry analysis give us in
addition the ellipticity, position angle and centers of ellipses as a
function of radius.  Fourier coefficients describing deviations from
pure ellipses yield information about the galaxy sub-structures (see
Falomo et al 1999).

\section{Results}
  For 64 sources the redshift is known while for three only
lower limits exist based on intervening absorption lines.  All the
objects with $z$ $< 0.5$ but two (OJ 287 and 0954+658) are resolved
in the HST images.  The detection of the host galaxy of OJ 287 in a GO HST image 
was claimed by Yanny et al (1997) but this is disputed by Sillanp\"{a}\"{a} et al. 
(this conference), while  for  second source
the redshift of is dubious.  We detected the host galaxy for
61 objects, 51 of which with known redshift. Two additional objects 
(0446+449 and 0525+713) are also well resolved but their classification as BL Lacs 
is uncertain (see Scarpa et al 1999). In the following we
concentrate on the 51 resolved BL Lacs of known $z$. 
Results for the unresolved sources as
well as for peculiar objects are given in Scarpa et al (1998,1999).

For all these objects the luminosity profile is well represented by an
elliptical galaxy model (convolved with the PSF) plus the contribution
of an unresolved nuclear source. In no case a disc dominated (spiral)
galaxy model gave a significantly better fit to the data. The superior
spatial resolution of HST compared with ground based data also helps to
detect possible high contrast structures (e.g. spiral arms) in the host
galaxy.  In none of the resolved objects such structures have been
detected, supporting the elliptical morphology classification.
\begin{figure}[h]
\plotfiddle{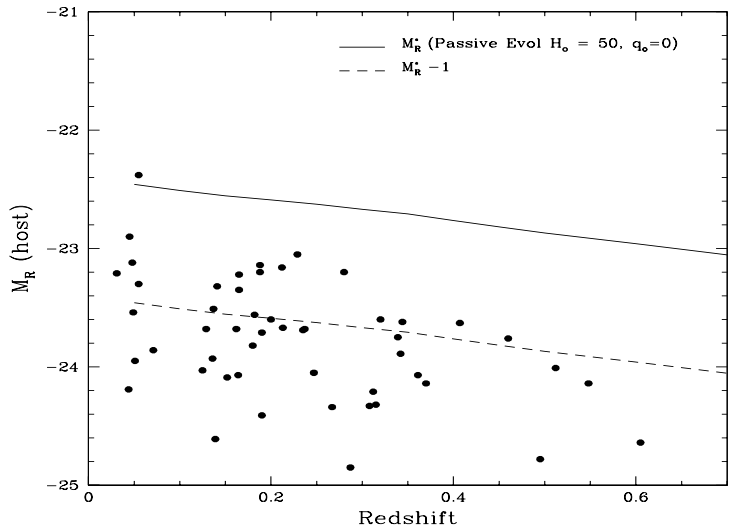}{6.5cm}{0}{50}{40}{-160}{-90}
\caption{Absolute magnitude of host galaxies of BL Lacs vs redshift. The solid line represents the luminosity of M$_R^*$ through passive 
evolution by Bressan et al 1994. Dashed line is the same for M$_R^*$ --1.}
\label{fig-1}
\end{figure}
The point source is always well centered onto the image of the
galaxy within an accuracy of 0.05 arcsec.  This argues against
microlensing playing an important role in BL~Lac objects,  
since in case of lensing by a foreground galaxy an off-centering of
the nucleus with respect to the surrounding nebulosity would be
observed (Ostriker \& Vietri 1985).

Using the results from the fitted profiles we have determined the
luminosity of the host galaxies for each source.  This is derived
integrating the fitted elliptical model to infinity (yielding total
magnitude {\it a la} de Vaucouleurs).  K-correction was applied using
a spectrum of a standard elliptical convolved with observed passband
and galactic extinction was taken into account using the Bell Lab
Survey of neutral hydrogen N$_H$ converted to E$_B-V$ (Stark et al
1992; Shull \& Van Steenberg 1985).  H$_0$ = 50~km~s$^{-1}$~kpc$^{-1}$
and q$_0 = 0$ are adopted.

In Figure 1 we plot the absolute magnitude of the host galaxy
M$_R$(host) as a function of the redshift.  For comparison we show the
expected luminosity of a M$_R^*$ elliptical with the passive evolution
model by Bressan et al 1994.  With only one exception all measured
hosts are brighter than M$_R^*$.  It is noticeable that the points
follow the expected trend of galaxy evolution.  The mean value of the
absolute magnitude of the hosts corrected for evolution
($<$M$_R$(host)$>$ = -23.6 $\pm$ 0.4) is $\sim$ 1 mag brighter than
M$_R^*$.  This value is consistent with previous findings from ground
based observations (Falomo 1996; Wurtz, Stocke \& Yee 1996) and
robustly confirms that host galaxies of BL Lac are luminous
ellipticals, on average brighter than L$^*$ but not as bright as
brightest cluster members.

\begin{figure}
\plotfiddle{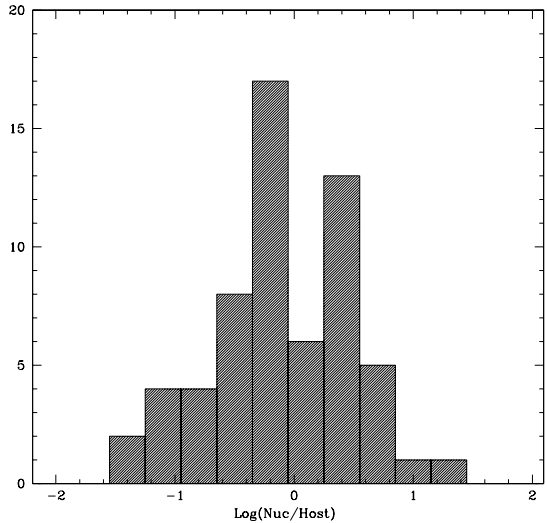}{5cm}{0}{40}{40}{-150}{-80}
\caption{Distribution of the nuclear/host luminosity.} \label{fig-2}
\end{figure}


The ratio of the nuclear over host luminosity for the resolved sources
is reported in Figure 2. The distribution peaks at $\sim$ 1 ($<$
Log[Nuc/Host]$>$ = -0.1).  On average therefore in the R band the
nuclear source has a luminosity similar to that of the galaxy.  The
nucleus/host luminosity ratio ranges from 0.1 (typical for radio
galaxies) to 10 (typical of quasars).

We show in Figure 3 the luminosity of the host versus that of the
nuclear source. While the spread of the magnitude of the host galaxies
is $\sim$ 1.5, the nuclear sources span over 5 magnitudes. We note
there is some deficit of host galaxies fainter than M$_R$(host) =
--23.5 objects with the nucleus brighter than M$_R$(nuc) =
--23.0. Since host galaxies in this region should be detectable this
suggests a tendency for the brightest galaxies to host the most
luminous nuclei.  However, because a number of sources have not been
resolved this point should be furthermore investigated including upper
limits of the host galaxies.

\begin{figure}
\plotfiddle{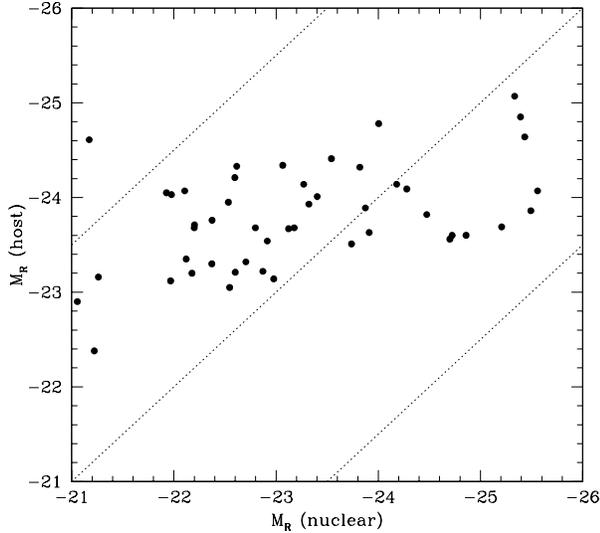}{5cm}{0}{40}{40}{-150}{-80}
\caption{M$_R$(nuc) versus M$_R$(host) of BL Lacs. The lines represent the 
loci of host/nucleus   
luminosity ratio corresponding to  0.1, 1, and 10 (bottom to top). } \label{fig-3}
\end{figure}

The overall morphology of host galaxies of BL Lacs is little perturbed
and appears, in this respect, different from the case of radio loud
quasars and powerful radio galaxies (e.g. Hutchings and Neff 1992; 
Smith and Heckman 1989)

In spite of the smoothness of host galaxy for many objects we have
detected close companions, consistently with previous suggestions
(Falomo 1996).  Two of them are very close ($< 0.5$ arcsec) to the
nucleus (see Scarpa et al 1998) and are extremely difficult to detect
from the ground.  Further scrutiny of these features is needed to
understand their nature and the role for the BL Lac phenomenon.

\begin{figure}[ht]
\plotfiddle{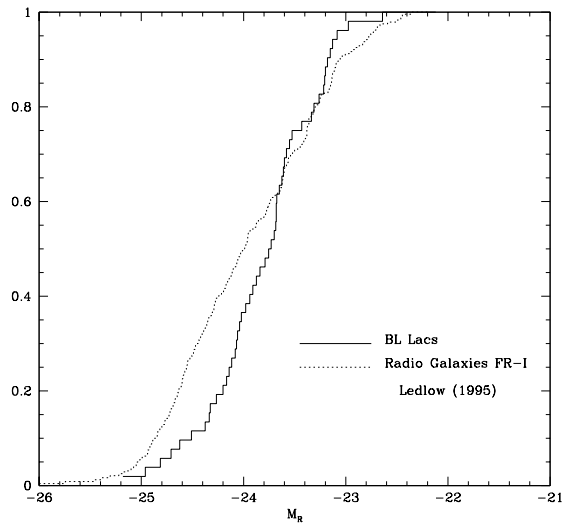}{6.5cm}{0}{40}{40}{-160}{-90}
\caption{Cumulative distributions of absolute magnitude of BL Lacs hosts 
({\it solid line}) and radio galaxies ({\it dotted line} Ledlow \& Owen).} \label{fig-4}
\end{figure}

Comparison with optical properties of radio galaxies is limited by the
lack of a large homogeneous data set in a wide range of redshifts. 
Moreover to perform an unbiased comparison one has to take into
account systematic effects like different observed spectral band,
K-correction, reddening and definition of galaxy magnitude.  Optical
studies of radio galaxies often use isophotal magnitudes and no
correction is done for possible presence of the nuclear sources.  This
translates into a spread of properties comparing different samples of
radio galaxies (Govoni et al 1999).

We have compared our distribution of host galaxy luminosity with that
of radio galaxies by Ledlow \& Owen (1995).  This is largest homogeneous
dataset of optical photometry of radio galaxies (mainly FR I radio
galaxies in Abell clusters).  The two cumulative distributions are
compared in Figure 4.  For M$_R >$ --23.5 the two distributions are
similar while at higher luminosities a deficit of BL Lac hosts with
respect to radio galaxies is apparent.  
This can be related to the fact that FR I radio galaxies
studied by Ledlow \& Owen are members of clusters.


\end{document}